\documentclass{article}

\input{tcilatex}

\begin{document}

{\footnotesize {Mathematical Problems of Computer Science {\small 25, 2006,
5--8.}}}\bigskip

\bigskip

\begin{center}
{\Large \textbf{Interval Edge Colourings of Complete Graphs and ${\large n}$%
-cubes}}

{\normalsize Petros A. Petrosyan}

{\small Institute for Informatics and Automation Problems (IIAP) of NAS of RA%
}

{\small e-mail: pet\_petros@yahoo.com}

\bigskip

\textbf{Abstract}
\end{center}

For complete graphs and n-cubes bounds are found for the possible number of
colours in an interval edge colourings.

\bigskip

Let $G=(V,E)$ be an undirected graph without loops and multiple edges [1], $%
V(G)$ and $E(G)$ be the sets of vertices and edges of $G$, respectively. The
degree of a vertex $x\in V(G)$ is denoted by $d_{G}(x)$, the maximum degree
of a vertex of $G$-by $\Delta (G)$, and the chromatic index of $G$-by $\chi
^{\prime }(G)$ and the diameter of $G$-by $d(G)$. A graph is regular, if all
its vertices have the same degree. If $\alpha $ is a proper edge colouring
[2] of the graph $G$, then $\alpha (e)$ denotes the colour of an edge $e\in
E(G)$ in the colouring $\alpha $. For a proper edge colouring $\alpha $ of
the graph $G$ and for any $x\in V(G)$ we denote by $S(x,\alpha )$ the set of
colours of edges incident with $x$.

An interval [3] $t-$colouring of $G$ is a proper colouring of edges of $G$
with colours $1,2,\ldots ,t$ such that at least one edge of $G$\ is coloured
by colour $i,1\leq i\leq t,$ and the edges incident with each vertex $x\in
V(G)$ are coloured by $d_{G}(x)$ consecutive colours.

A graph $G$ is interval-colourable if there is $t\geq 1$ for which $G$ has
an interval $t-$colouring. The set of all interval-colourable graphs is
denoted by $\mathcal{N}$ [4].

For $G\in \mathcal{N}$ we denote by $w(G)$ and $W(G)$, respectively, the
least and greatest value of $t$, for which $G$ has an interval $t-$colouring.

The problem of deciding whether or not a bipartite graph belongs to $%
\mathcal{N}$ was shown in [5] to be $NP$-complete [6,7].

It was proved [4] that if $G$ has no triangle and $G\in \mathcal{N}$ then $%
W(G)\leq \left\vert V(G)\right\vert -1$. It follows from here that if $G$ is
bipartite and $G\in \mathcal{N}$ then $W(G)\leq \left\vert V(G)\right\vert
-1 $.

\textbf{Theorem 1 }[8]. If $G$ is a bipartite graph and $G\in \mathcal{N}$,
then $W(G)\leq d(G)(\Delta (G)-1)+1$.

For graphs which can contain a triangle the following results hold:

\textbf{Theorem 2 }[4]. If $G\in \mathcal{N}$ is a graph with nonempty set
of edges then $W(G)\leq 2\left\vert V(G)\right\vert -3$.

\textbf{Theorem 3 }[9]. If $G\in \mathcal{N}$ and $\left\vert
V(G)\right\vert \geq 3$ then $W(G)\leq 2\left\vert V(G)\right\vert -4$.

In [4] there was proved the following

\textbf{Theorem 4}. Let $G$ be a regular graph.

1) $G\in \mathcal{N}$ \ iff \ $\chi ^{\prime }(G)=\Delta (G)$.

2) If $G\in \mathcal{N}$ and $\Delta (G)\leq t\leq W(G)$ then $G$ has an
interval $t-$colouring.

From Theorem 4 and the result of\textbf{\ }[10] it follows that the problem
\textquotedblleft Does a given regular graph belongs to the set $\mathcal{N}$
or not?\textquotedblright\ is $NP$-complete.

In this paper interval edge colourings of complete graphs and $n$-cubes are
investigated. Non-defined conceptions and notations can be found in
[1,2,4,8].

From the results of [11] and Theorem 4 it follows that for any odd $p$ $\ \
K_{p}\notin \mathcal{N}$. It's easy to see that $\ \chi ^{\prime
}(K_{2n})=\Delta (K_{2n})=2n-1$ [1] and, therefore for any $\ n\in N$ $\ \
K_{2n}\in \mathcal{N}$ , $w(K_{2n})=2n-1$.

\textbf{Theorem 5 }[12]. For any \ $n\in N$ \ \ $W(K_{2n})\geq 3n-2$.

\textbf{Theorem 6}. Let $n=p2^{q}$, where $p$ is odd and $q$ is nonnegative
integer. Then $W(K_{2n})\geq 4n-2-p-q$.

\textbf{Proof}. Let's prove that for any $m\in N$ \ \ $W(K_{4m})-W(K_{2m})%
\geq 4m-1$.

Consider a graph $K_{4m}$ with $V(K_{4m})=\left\{
x_{1},x_{2},...,x_{4m}\right\} $ and

$E(K_{4m})=\left\{ \left( x_{i},x_{j}\right) |\text{ \ }x_{i}\in V(K_{4m}),%
\text{ }x_{j}\in V(K_{4m}),\text{ }i<j\right\} $.

Let $G$ be a subgraph of the graph $K_{4m}$, induced by its vertices $%
x_{1},x_{2},...,x_{2m}$.\ Evidently $G$ is isomorphic to the graph $K_{2m}$
and, consequently, there exists an interval $W(K_{2m})-$colouring $\alpha $
of $G$.

Now we define an edge colouring $\beta $ of $K_{4m}$.

For $i=1,2,...,4m$ \ and $j=1,2,...,4m$, where $i\neq j$, we set:

\bigskip

$\beta ((x_{i},x_{j}))=\left\{ 
\begin{array}{c}
\alpha ((x_{i},x_{j}))\text{,\ }\ \ \ \ \ \ \ \ \ \ \ \ \ \ \ \ \ \ \ \ \ \
\ \ \ \ \ \ \ \ \ \ \ \ \ \ \ \ \text{if}\ 1\leq \ i\leq 2m,\text{ }1\leq
j\leq 2m;\  \\ 
\min S(x_{i},\alpha )+2m-1\text{,\ if}\ 1\leq i\leq 2m,\text{ }2m+1\leq
j\leq 4m,\text{ }i=j-2m; \\ 
\alpha ((x_{i},x_{j-2m}))+2m\text{,\ \ if}\ 1\leq i\leq 2m,\text{ }2m+1\leq
j\leq 4m,\text{ }i\neq j-2m;\  \\ 
\alpha ((x_{i-2m},x_{j-2m}))+4m-1\text{,\ if}\ 2m+1\leq i\leq 4m,\text{ }%
2m+1\leq j\leq 4m.%
\end{array}%
\right. $

\bigskip

It is not difficult to see that $\ \beta $ \ is an interval $%
(W(K_{2m})+4m-1)-$colouring of $K_{4m}$.

Now we can conclude:

\begin{center}
$\QDATOP{W(K_{p2^{q+1}})\geq W(K_{p2^{q}})+p2^{q+1}-1}{W(K_{p2^{q}})\geq
W(K_{p2^{q-1}})+p2^{q}-1}$

\bigskip $\QDATOP{\cdots \cdots \cdots \cdots \cdots \cdots \cdots \cdots
\cdots \cdots \cdots \cdots }{W(K_{p2^{2}})\geq W(K_{p2})+p2^{2}-1}$
\end{center}

Adding these inequalities we obtain

\begin{center}
$W(K_{2n})\geq W(K_{2p})+p\sum\limits_{i=2}^{q+1}2^{i}-q$.
\end{center}

Now, using the result of Theorem 5, we have

\begin{center}
$W(K_{2n})\geq
3p-2-q+p\sum\limits_{i=2}^{q+1}2^{i}=3p-2-q+4p(2^{q}-1)=4n-2-p-q$.
\end{center}

The proof is complete.

\textbf{Corollary 1}. Let $n=p2^{q}$, where $p$ is odd and $q$ is
nonnegative integer. If $\ 2n-1\leq t\leq 4n-2-p-q$ \ then there exists an
interval $t-$colouring of $K_{2n}$.

\textbf{Lemma}. For any $\ n\in N$ $\ \ \ Q_{n}\in \mathcal{N}$ \ and \ $%
w(Q_{n})=n$.

\textbf{Proof}. As for any $\ n\in N$ $\ \ \ Q_{n}$ is a regular bipartite
graph then $\chi ^{\prime }(Q_{n})=\Delta (Q_{n})=n$ and, from Theorem 4, $%
Q_{n}\in \mathcal{N}$\ , $w(Q_{n})=n$.

\textbf{Theorem 7}. For any \ $n\in N$ \ \ $W(Q_{n})\geq \dfrac{n(n+1)}{2}$.

\textbf{Proof}. Let's prove that for $n\geq 2$\ \ $W(Q_{n})-W(Q_{n-1})\geq n$%
.

Evidently, $Q_{n}=K_{2}\times Q_{n-1}$, therefore there are two subgraphs $%
Q_{n-1}^{(1)}$ and \ $Q_{n-1}^{(2)}$ of $Q_{n}$, which satisfy conditions:

\begin{center}
$V(Q_{n-1}^{(1)})\cap V(Q_{n-1}^{(2)})=\emptyset $,

$Q_{n-1}^{(i)}$is isomorphic to $Q_{n-1}$, $i=1,2$.
\end{center}

It follows from Lemma that for\ $i=1,2$ \ $Q_{n-1}^{(i)}\in \mathcal{N}$.

Evidently $Q_{n-1}^{(1)}$ is isomorphic to $Q_{n-1}^{(2)}$, therefore there
exists a bijection $f$ $:V(Q_{n-1}^{(1)})\longrightarrow V(Q_{n-1}^{(2)})$
such that $\left( x,y\right) \in E(Q_{n-1}^{(1)})$ iff $\left( f\left(
x\right) ,f\left( y\right) \right) \in E(Q_{n-1}^{(2)})$. Let $\alpha $ be
an interval $W(Q_{n-1}^{(1)})-$colouring of the graph $Q_{n-1}^{(1)}$.

Let's define an edge colouring $\beta $ of the graph $Q_{n-1}^{(2)}$ in the
following way: for every edge $(u,v)\in E(Q_{n-1}^{(2)})$\ \ \ $\beta
((u,v))=\alpha ((f^{-1}(u),f^{-1}(v)))+n$.

Now we define an edge colouring $\gamma $ of the graph $Q_{n}$.

For every edge $\left( x,y\right) \in E(Q_{n})$

\bigskip

$\gamma ((x,y))=\left\{ 
\begin{array}{c}
\alpha ((x,y))\text{,\ \ \ \ \ \ \ \ \ \ \ \ \ \ \ \ \ \ \ \ \ \ \ \ \ \ \ }%
\ \ \ \text{if }x\in V(Q_{n-1}^{(1)}),y\in V(Q_{n-1}^{(1)}); \\ 
\min S(x,\alpha )+n-1\text{, \ if}\ x\in V(Q_{n-1}^{(1)}),y\in
V(Q_{n-1}^{(2)}),\text{ }y=f(x); \\ 
\beta ((x,y))\text{,\ \ \ \ \ \ \ \ \ \ \ \ \ \ \ \ \ \ \ \ \ \ \ \ }\ \ \ \
\ \text{if}\ \ x\in V(Q_{n-1}^{(2)}),\text{\ }y\in V(Q_{n-1}^{(2)})\text{.}\ 
\end{array}%
\right. $

\bigskip

It is not difficult to see that $\gamma $ \ is an interval $(W(Q_{n-1})+n)-$%
colouring of $Q_{n}$.

For $n\geq 2$ we have

\begin{center}
$\QDATOP{W(Q_{n})\geq W(Q_{n-1})+n}{W(Q_{n-1})\geq W(Q_{n-2})+n-1}$

$\cdots \cdots \cdots \cdots \cdots \cdots \cdots \cdots \cdots $

$W(Q_{2})\geq W(Q_{1})+2$
\end{center}

Adding these inequalities we obtain \ $W(Q_{n})\geq \dfrac{n(n+1)}{2}$.

The proof is complete.

\textbf{Corollary 2}. If $\ \ n\leq t\leq $ $\dfrac{n(n+1)}{2}$ \ then $%
Q_{n} $ \ has an interval $t-$colouring.

$\bigskip $

\begin{center}
\bigskip
\end{center}

\end{document}